\documentclass[prd,onecolumn,notitlepage,showpacs,preprintnumbers,amsmath,amssymb,nofootinbib,APS,10pt,superscriptaddress]{revtex4-1}

\usepackage{dcolumn}
\usepackage{bm}
\usepackage{xcolor}
\usepackage[applemac]{inputenc}
\usepackage[spanish,english]{babel}
\usepackage{amsmath,amssymb,amsfonts,latexsym,cancel}
\usepackage{graphicx}
\usepackage{color}
\usepackage{soul}
\usepackage{ulem}
\usepackage{hyperref}
\usepackage{amsmath}
\usepackage{slashed}
\usepackage{float}

\allowdisplaybreaks

\newcommand{\be}{\begin{equation}}
\newcommand{\ee}{\end{equation}}
\newcommand{\bea}{\begin{eqnarray}}
\newcommand{\eea}{\end{eqnarray}}

\begin{document}

\title{{\bf $R$-summed form of adiabatic expansions in curved spacetime}}

\author{Antonio Ferreiro}\email{antonio.ferreiro@ific.uv.es}
\author{Jose Navarro-Salas}\email{jnavarro@ific.uv.es}
\author{Silvia Pla}\email{silvia.pla@uv.es}

\affiliation{Departamento de Fisica Teorica and IFIC, Centro Mixto Universidad de Valencia-CSIC. Facultad de Fisica, Universidad de Valencia, Burjassot-46100, Valencia, Spain.}

\begin{abstract}

The Feynman propagator in curved spacetime admits an asymptotic (Schwinger-DeWitt) series expansion in derivatives of the metric. Remarkably, all terms in the series containing the Ricci scalar $R$ can be summed exactly.  
We show that this (nonperturbative) property of the Schwinger-DeWitt series has a natural and equivalent counterpart in the adiabatic (Parker-Fulling) series expansion of the scalar modes in an homogeneous cosmological spacetime. The equivalence  between both $R$-summed adiabatic expansions can be further  extended  when a background scalar field is also present. \\

\end{abstract}


\date{\today}
\maketitle

\section{Introduction}

One of the most useful tools in the theory of quantized fields in curved spacetime \cite{parker-toms, fulling, birrell-davies} and semiclassical gravity \cite{hu-verdaguer} is the Schwinger-DeWitt (SDW) adiabatic  (proper-time) expansion of the Feynman propagator \cite{DeWittbook}. It consists in an expansion in  number of derivatives of the metric with a fixed leading term. This expansion is of utmost importance in the renormalization of  expectation values of the stress-energy tensor. It also plays a fundamental role in the evaluation of the effective action. 
The SDW expansion identifies the ultraviolet (UV) divergences of Green's functions in a generic spacetime and it can be accompanied  with the point-splitting  technique  \cite{dewitt75, christensen76} to renormalize expectation values of observables such as the  stress-energy tensor. The SDW representation of the Feynman two-point function for a scalar field can be regarded as a special case of the Hadamard expansion, corresponding to a particular choice of the undetermined biscalar coefficient in the Hadamard representation \cite{waldbook}. The SDW expansion can also be  rederived from the local momentum-space representation introduced by Bunch and Parker \cite{bunch-parker}. In this context, Bekenstein and Parker \cite{Bekenstein-Parker} obtained  an approximated form for the propagator (the Gaussian approximation) involving, in the coincidence limit, an exponential of the scalar curvature $R$. 
 Remarkably, it has been conjectured by Parker and Toms \cite{Parker-Toms85} and proved to all orders by Jack and Parker \cite{Jack-Parker}  that  this nonperturbative exponential  factor $\exp[-is(\xi-1/6)R]$ is indeed the  sum of all terms containing $R$ in the adiabatic proper-time series. This result has major physical consequences to account for the effective dynamics of the Universe and the observed cosmological acceleration. By integrating out the quantum fluctuations of an ultra-low-mass scalar field the effective gravitational dynamics provides negative pressure  to suddenly accelerate the Universe, without the need of an underlaying cosmological constant \cite{parker-raval,parker-ravalA, parker-vanzella} (see also Ref. \cite{caldwell-komp-parker-vanzella}).  This approach can also alleviate \cite{Melchiorri} the increasing $H_0$ tension of the standard  cosmological model. Other  physical applications are reported in Refs. \cite{Tommi, prl, JHEP}.\\ 

Within the cosmological context, and for  Friedmann-Lemaitre-Robertson-Walker (FLRW) spacetimes, it is convenient to  regard the Feynman Green's function as a sum in modes. The modes themselves admit an adiabatic expansion, also with a fixed leading term, in  number of derivatives of the expansion factor $a(t)$. The modes of a scalar field have a natural adiabatic expansion, which generalize the so called Wentzel-Kramers-Brillouin (WKB) approximation. This adiabatic expansion can be exploited to compute the renormalized expectation values of the stress-energy tensor, as first proposed and studied by Parker and Fulling (PF) \cite{parker-fulling74} (for a historical account, see Ref. \cite{parker12}). This adiabatic method  identifies  the UV subtracting terms directly in momentum space.  One advantage of the PF adiabatic expansion relies on the systematics of the algorithm to determine  arbitrary higher-order adiabatic terms. Furthermore, it is also a very efficient method of renormalization in homogeneous cosmological spacetimes \cite{adiabatic1,adiabatic2,adiabatic3,adiabatic5,adiabatic6,adiabatic7,adiabatic8},   especially  in studies in which numerical computations are finally required. The method has been extended to deal with Dirac fields \cite{Dirac1,Dirac1a,Dirac2,Dirac3,Dirac3a,Dirac4} and with scalar \cite{adiabatic4,adiabatic4b} and electromagnetic backgrounds \cite{em0,em1,em2,em3,em4,em5}. In FLRW backgrounds both adiabatic schemes of renormalization (PF and SDW) can be applied, and these methods can be shown to be equivalent \cite{delRio-Navarro15, Birrell78, Anderson-Parker87}. See Ref. {\cite{winstanley} for a discussion on the equivalence among different  renormalization schemes.  \\


The aim of this work is to show that the $R$-summed form of the adiabatic Schwinger-DeWitt expansion of the propagator has an equivalent  counterpart in the adiabatic Parker-Fulling expansion of the field modes. 
We will also show that this result is naturally extended  when a background scalar field is present, as happens in the single field models of inflation \cite{Liddle-Lyth00, Dodelson03}. \\ 

The paper is organized as follows. In Sec. \ref{sec2} we  briefly introduce the  (Schwinger-DeWitt) proper-time expansion of the Feynman propagator  and the Parker-Fulling  adiabatic expansion of the  field modes  on a FLRW spacetime. We also describe the equivalence between both adiabatic expansions. In Sec. \ref{sec3}  we introduce the $R$-summed form of the SDW expansion and propose a new $R$-summed form of the traditional adiabatic WKB-type expansion  of  the field modes in a FLRW spacetime.  We provide strong evidence for the equivalence between both ($R$-summed) expansions. In Sec.  \ref{sec4}   we generalize the previous result by also including a  classical background scalar field with a Yukawa-type coupling  to the quantized scalar field. 
As a simple byproduct of our analysis we  also give the effective Lagrangian  
induced by quantum fluctuations of the quantized scalar field. 
 Finally, in Sec. \ref{sec5}  we summarize our main conclusions. 

\section{Schwinger-DeWitt and  Parker-Fulling adiabatic expansions}\label{sec2}
\subsection{Schwinger-DeWitt adiabatic expansion}
Let us consider a quantized scalar field $\phi$ on a general  smooth four-dimensional  spacetime. The associate  Feynman propagator $G(x, x')=-i \langle 0| T \phi(x) \phi(x') |0\rangle$ satisfies the equation
\be (\Box_x + m^2 + \xi R) G(x, x') = -|g(x)|^{-1/2} \delta (x-x') \label{green1} \ , \ee
where $\xi$ parametrizes the coupling to the scalar curvature. We follow the convention and notation given in Ref. \cite{parker-toms}. 
To implement the renormalization program it is very useful to construct an adiabatic  expansion of $G(x,x')$  in terms of the number of derivatives of the background metric. This is the basic idea of the 
SDW expansion \cite{DeWittbook}. 
To obtain the desired expansion, one writes the propagator in terms of the proper-time form
\be \label{GFs} G(x, x') = -i \int_0^\infty ds \ e^{ -im^2 s} \langle x, s |  x', 0\rangle \ee 
where $m^2$ is understood to have an infinitesimal negative imaginary part $-i\epsilon$. The kernel $\langle x, s |  x', 0\rangle$ satisfies the Schr\"odinger-type equation
\be \label{hk}i\frac{\partial }{\partial s} \langle x, s |  x', 0\rangle = (\Box_x + \xi R)\langle x, s |  x', 0\rangle \ , \ee
with the boundary condition $\langle x, s |  x', 0\rangle \sim |g(x)|^{-1/2}\delta(x-x')$ as $s\to 0$. Equation (\ref{hk}) implies that, by iteration,  $\langle x, s |  x', 0\rangle$ can be further expanded in powers of the proper-time parameter.{\footnote{In spacetimes with boundaries and singularities there are  additional terms (see for instance Ref. \cite{Vassilevich}). }} This can be made explicit by introducing a function $F(x, x'; is)$ defined by  
the relation
{\be \label{hks}\langle x, s |  x', 0\rangle =  i\frac{\Delta^{1/2} (x, x')}{(4\pi)^2(is)^2} e^{\frac{\sigma(x, x')}{2is}} \ F(x, x'; is) \ , \ee 
where $\Delta(x, x')$ is the Van Vleck-Morette determinant and $\sigma(x, x')$ is  the proper distance along  the geodesic from $x'$ to $x$.
The asymptotic expansion of the function $F(x, x';is)$ is
\be  \label{F1}F(x, x'; s) \sim a_0(x, x') + a_1(x, x') (is) + a_2(x, x') (is)^2 + \cdots  \ , \ee
where the first coefficients $a_n(x, x')$ are given, in the coincidence limit $x \to x'$ (see Ref. \cite{DeWittbook}): 

\bea &&a_0(x)=1 \ , ~~~~a_1(x)=-\bar \xi R \ , \\ 
&&a_2(x)=\frac{1}{180}R_{\alpha\beta\gamma\delta}R^{\alpha\beta\gamma\delta}-\frac{1}{180}R^{\alpha\beta}R_{\alpha\beta}-\frac{1}{6}\left(\frac{1}{5}-\xi\right)\Box R+\frac{1}{2}{\bar {\xi}}^2 R^2\label{Coef1}  
\ , \eea
and $\bar\xi\equiv \xi- \frac{1}{6}$. Higher-order coefficients $a_n$ have been calculated in Refs. \cite{sakai, gilkey, avramidi, avramidi2}. Hence, the SDW expansion, at a given adiabatic order $2n$,  takes the form
\be
\label{Gn} ^{(2n)}G_{SDW}(x, x') =  \frac{\Delta^{1/2} (x, x')}{(4\pi)^2} \int_0^\infty\frac{ds}{(is)^2} e^{ -im^2 s} e^{\frac{\sigma(x, x')}{2is}} \sum_{j=0}^n a_j(x,x') (is)^j \ . 
\ee
 We recall here that the coefficient $a_j$ is of adiabatic order $2j$.  In four spacetime dimensions, and for arbitrary $\xi$, the first two terms in \eqref{F1} make \eqref{GFs} divergent   in the UV limit, namely, when $s\to 0$ and $\sigma=0$. For instance, the first two leading terms in the adiabatic expansion are, after performing the $ds$ integral, 
\bea \label{DSG4}
^{(2)} G_{SDW}(x,x') & = & -i\frac{|g(x)|^{-1/4}}{4\pi^2}\left[  \frac{m}{\sqrt{-2\sigma}} K_1(m\sqrt{-2\sigma})+\frac{a_1(x, x')}{2}K_0 (m\sqrt{-2\sigma}) \right] \label{result} \ , \ \ \ \ \ \ 
\eea
where  $K$ are the modified Bessel functions of second kind. The factor $|g(x)|^{-1/4}$ in the above expression is evaluated in Riemann normal coordinates with origin at $x'$ \cite{delRio-Navarro15}.  Higher-order terms do not involve any UV divergences for the two-point function. However, the fourth adiabatic order term, $a_2$, is necessary to tame the logarithmic divergences of the stress-energy tensor and the effective action \cite{dewitt75, christensen76} (see also Refs. \cite{parker-toms, birrell-davies}).  \\ 

There are two issues to note on this expansion. First, one can extend the series to an arbitrary order, 
although only the first few  terms are analytically manageable. Second, any higher-order term contains only polynomial terms of the curvature, to such an extent that nonlocal effects are not present at any given adiabatic order. The later was taken into consideration in Refs. \cite{Parker-Toms85,Jack-Parker} by proposing a refined expansion which we will describe in the next section. Nevertheless, we will first briefly review to adiabatic Parker-Fulling expansion and the equivalence between this and the SDW expansion in FLRW spacetimes. This is required in order to extend the Parker-Fulling expansion and overcome the former issue.  \\

\subsection{Parker-Fulling adiabatic expansion} \label{PFadiabaticexp}


Let us assume for simplicity a spatially flat metric of the form $ds^2=dt^2 -a^2(t)d\vec{x}^2$. The scalar field satisfies the equation
\bea
(\square+m^2+\xi R)\phi=0\label{eqm}
\ , \eea
where $R=6(\dot a^2/a^2+\ddot a/a)$. The quantized field is expanded in Fourier modes as
\be \phi(x)= \frac{1}{\sqrt{2(2 \pi a^3)}}\int d^3 \vec{k} [A_{\vec{k}} f_{\vec{k}}(x)+A_{\vec{k}}^{\dagger}f_{\vec{k}}^{*}(x) ] \ , \label{phisolution} \ee
where $f_{\vec{k}}(x) = e^{i\vec{k}\vec{x}}h_k(t)$ and
 $A_{\vec{k}}^{\dagger}$ and $A_{\vec{k}}$ are the usual creation and annihilation operators. Substituting \eqref{phisolution} into (\ref{eqm})  we find $\ddot h_k+ \left[ \omega^2+\sigma \right] h_k=0$,
where $\sigma=(6\xi- \frac{3}{4})(\frac{\dot{a}^2}{a^2})+(6\xi-\frac{3}{2})(\frac{\ddot{a}}{a})$ and $\omega=\sqrt{\frac{k^2}{a^2}+m^2}$.
The adiabatic expansion for the scalar field modes is based on the usual WKB ansatz \cite{parker-toms, fulling, birrell-davies}
\be \label{WKB}h_k(t)= \frac{1}{\sqrt{W_k(t)}} e^{-i\int^t W_k (t') dt'} \ , \hspace{0.5cm}   W_k(t) = \omega^{(0)} + \omega^{(1)} + \omega^{(2)} + \cdots \ , \ee
where the adiabatic order  is based on the number of derivatives of the expansion factor $a(t)$. The function $W_k(t)$ obeys the differential equation
\be   W_k^2=  \omega^2+ \sigma+\frac{3}{4}\frac{\dot{ W}_k^2}{ W^2_k} -\frac 12\frac{\ddot{ W}_k}{W_k}\  \label{equWk2} \ . \ee
If we now fix the leading term as 
$\omega^{(0)}= \omega $, one can substitute  the ansatz into Eq. (\ref{equWk2}), and solve order by order to obtain recursively the different terms of the expansion:
\bea \label{omegan}
&&\omega^{(1)}=\omega^{(3)}=0 \nonumber \\
&&\omega^{(2)}=\frac{1}{2\omega^3}\Big\{\sigma \omega^2+\frac{3}{4}\dot \omega ^2-\frac 12 \omega \ddot \omega \Big\} \nonumber \\
&&\omega^{(4)}=\frac{1}{2\omega^3}\Big\{2\sigma \omega \omega^{(2)}-5\omega^2(\omega^{(2)})^2+\frac{3}{2}\dot \omega\dot \omega^{(2)}-\frac 12 (\omega \ddot \omega^{(2)}+ \omega^{(2)} \ddot \omega) \Big\} \label{w4}\ .
\eea
Note that, in this expansion, the coefficients of odd adiabatic order, namely $\omega^{(2n+1)}$, are always zero. From the mode expansion, we can expand any observable at any fixed adiabatic order. For the two-point function at the coincident limit  $G(x,x)\sim \int dk k^2 W^{-1}_k$, we have
\bea
\label{gpf1}^{(2n)}G_{PF}(x,x)= \frac{1}{4\pi^2 a^3(t)}\int_0^\infty dk k^2 \left \{  \omega^{-1} + (W^{-1})^{(2)}+ (W^{-1})^{(4)}+... + (W^{-1})^{(2n)}\right\}
\ , \eea
where the first  terms are 
 \bea &&(W^{-1})^{(2)}= \frac{m^2\dot a^2}{2a^2\omega^5} + \frac{m^2\ddot a}{4a\omega^5} -\frac{5m^4\dot a^2}{8a^2\omega^7}-\frac{\bar \xi R}{2\omega^3}\label{W-1(2)} \\ && (W^{-1})^{(4)}= -\frac{\omega^{(4)}}{\omega^2}+\frac{\left(\omega^{(2)}\right)^2}{\omega^3} \ .\eea

Just as the SDW expansion, only the first two terms in \eqref{gpf1} are divergent, in such a way that it serves to isolate all the ultraviolet divergences of the propagator. 
 More precisely, we have 
\be \label{adiabaticresult} ^{(2)}G_{PF}(x,x)=    \frac{R}{288 \pi^2}  +  \frac{1}{4 \pi^2 a^3}  \int_0^{\infty} d k k^2 \,  \left[ \frac{1}{\omega} - \frac{\bar\xi R}{2\omega^3}\right]   \ . \ee 
After subtracting the divergences, one gets  a finite result. This mechanism  can also used to renormalize the expectation values of the stress-energy tensor. The overall procedure is traditionally known as the adiabatic regularization  method \cite{parker-toms, parker-fulling74}.  Even though we have written \eqref{equWk2} in a compact form, we can further expand this expression and obtain an analytic expression for $\omega^{(2n)}$ in terms of the lower adiabatic orders (see for instance Ref. \cite{delRio-Navarro15}). \\

\subsection{Comparison between the Schwinger-DeWitt and Parker-Fulling adiabatic expansions}
To  compare both adiabatic expansions, we have to restrict the Schwinger-DeWitt expansion of the Feynman propagator to the (spatially flat) FLRW universe considered above. Moreover, it is natural to compare the expansion of the two-point function $G_F(x, x')$ at the coincident limit $x=x'$. The comparison is highly nontrivial since in the SDW formalism the coincidence limit is defined in terms of the geodesic distance with $\sigma \to 0$. We follow the analysis in Ref. \cite{delRio-Navarro15}. \\

 The zeroth-order contribution  $^{(0)} G_{SDW}(x,x)$ can be reexpressed as [here $x\equiv (t, \vec x)$ and $x'\equiv (t, \vec x')$]
\bea
\lim_{x\to x\, '}    \frac{|g(x)|^{-1/4} m }{(2\pi)^2 \sqrt{-2\sigma}}K_1(m\sqrt{-2\sigma}) & = & \frac{R}{288 \pi^2}+  \lim_{\Delta \vec x\to 0}   \frac{m }{4 \pi^2 a|\Delta \vec x|}K_1(m\, a|\Delta \vec x|) \nonumber \\
& = & \frac{R}{288 \pi^2}+  \lim_{\Delta \vec x\to 0}   \frac{1}{4 \pi^2 a^3}\int_0^{\infty} dk k^2 \frac{\sin(k |\Delta \vec x|)}{k |\Delta \vec x|} \frac{1}{\omega} \ ,
\eea
where we have used
\bea
\frac{1}{-2\sigma} & = & \frac{1}{a^2 \Delta \vec x^2}-\frac{\dot a^2}{12a^2}+O(\Delta \vec x^2) \ , \label{I1}
\eea
and 
\bea
|g(x)|^{-1/4} & = & 1-\left[ 2\frac{\dot a^2}{a^2}+\frac{\ddot a}{a} \right] \frac{\sigma}{6}+O(\sigma^{3/2})\ . \label{I2}
\eea
Similarly,  the second-order adiabatic contribution to $^{(2)} G_{SDW}(x,x)$ is found to be
\bea
 \lim_{x\to x\, '} \frac{|g(x)|^{-1/4}}{4\pi^2}\frac{a_1(x, x')}{2}K_0 (m\sqrt{-2\sigma})
  & = &   \lim_{\Delta \vec x\to 0} -\frac{1}{4\pi^2}\frac{\bar \xi R}{2}K_0 (m\, a|\Delta \vec x|) \nonumber \\
  & = &\lim_{\Delta \vec x\to 0} \frac{-1}{4 \pi^2 a^3}\int_0^{\infty} dk k^2 \frac{\sin(k |\Delta \vec x|)}{k |\Delta \vec x| } \frac{\bar \xi R}{2\omega^3} 
\ . \eea
Therefore, taking into account (\ref{adiabaticresult}), one can write
\be {^{(2)}}G_{PF}(x, x)= i {^{(2)}}G_{SDW}(x, x)=  \frac{1}{4 \pi^2 a^3}  \int_0^{\infty} d k k^2 \,  \left[ \frac{1}{(\frac{k^2}{a^2} + m^2)^{1/2}} - \frac{\bar \xi R(x)}{2(\frac{k^2}{a^2} +m^2)^{3/2}}\right] + \frac{R(x)}{288 \pi^2} 
\ . \ee
A detailed analysis can be found in Ref. \cite{delRio-Navarro15}. It was explicitly checked (up to and including the sixth adiabatic order) that the Parker-Fulling expansion of the  two-point function $G_{PF}(x,x)$ coincides with the corresponding Schwinger-DeWitt expansion of the two-point function at coincidence $G_{SDW}(x,x)$, that is,
\be \label{eGDSAd6}{^{(6)}}G_{PF}(x, x)= i {^{(6)}}G_{SDW}(x, x)=  \frac{1}{4 \pi^2 a^3}  \int_0^{\infty} d k k^2 \,  \left[ \frac{1}{(\frac{k^2}{a^2} + m^2)^{1/2}} - \frac{\bar \xi R(x)}{2(\frac{k^2}{a^2} +m^2)^{3/2}}\right] + \frac{R(x)}{288 \pi^2}  +\frac{a_2(x)}{16\pi^2 m^2} + \frac{a_3(x)}{16\pi^2m^4} \ . \ee
This provides enough evidence for the equivalence at any adiabatic order, 
\be
\label{equiv1}{^{(2n)}}G_{PF}(x, x)= i {^{(2n)}}G_{SDW}(x, x) \ .
\ee
In the next section, we will show that this equivalence can also be extended to the $R$-summed form of the SDW expansion given in Ref. \cite{Parker-Toms85, Jack-Parker}.

\section{$R$-summed form of the adiabatic expansions}\label{sec3}

As stressed in the Introduction, a very important result concerning the SDW adiabatic expansion is that the   expansion of the kernel  $\langle x, s |  x', 0\rangle$ of \eqref{GFs} can be rewritten in the form \cite{Parker-Toms85, Jack-Parker}
\be \label{DSR1} \langle x, s |  x', 0\rangle =  i\frac{\Delta^{1/2} (x, x')}{(4\pi)^2(is)^2}  e^{\frac{\sigma(x, x')}{2is}} e^{ -i\bar \xi R(x')s} \bar F (x, x'; is) \ , \ee 
 where $ \bar F (x, x'; is)$ is the new proper-time series 
\be \bar F (x, x'; is) = \sum_j (is)^j \bar a_j (x,x') \label{DSR}\ . \ee  
It has been proven that for general spacetimes in arbitrary dimensions, this expansion depends on $R$ only by the overall exponential factor  (it can contain nevertheless derivatives of the  scalar curvature).  In particular, for the first terms, we have (see Appendix \ref{Apan} for more details) 
 \bea && \bar a_0(x) =1, \ \ \ \ \bar a_1(x) = 0, \\
&& \bar a_2(x) = \frac{1}{180}R_{\alpha\beta\gamma\delta}R^{\alpha\beta\gamma\delta}-\frac{1}{180}R^{\alpha\beta}R_{\alpha\beta}-\frac{1}{6}\left(\frac15-\xi\right)\Box R \ . \eea
The $R$-summed form of the  SDW expansion takes the form 
\be
^{(2n)}\bar{G}_{SDW}(x, x') =  \frac{\Delta^{1/2} (x, x')}{(4\pi)^2} \int_0^\infty\frac{ds}{(is)^2} e^{ -i(m^2+\bar \xi R) s} e^{\frac{\sigma(x, x')}{2is}} \sum_{j=0}^n (is)^j \bar a_j (x,x') \ . 
\ee
 Expansion \eqref{DSR} indicates that there is a subset of the original  series that can be exactly summed and factorized out in the exponential term of \eqref{DSR1}. As we have already mentioned, this nonperturbative effect was further applied to study the observed cosmological acceleration \cite{parker-raval,parker-ravalA,parker-vanzella}. The main feature of this expansion is that no $R$ term appears explicitly in the $\bar a_n$ coefficients. We will inherit this characteristic for the Parker-Fulling expansion. The natural way of doing this is to include the same $R$-summed contribution of \eqref{DSR1} into the leading term of the adiabatic expansion, namely in $\omega^{-1}$. We note that the terms 
 \be 
\left[ \frac{1}{(\frac{k^2}{a^2} + m^2)^{1/2}} - \frac{\bar \xi R}{2(\frac{k^2}{a^2} +m^2)^{3/2}}\right] \ee
in the momentum integral in (\ref{eGDSAd6}) can be regarded as the leading terms in the expansion of 
\be \frac{1}{(\frac{k^2}{a^2} + m^2+ \bar \xi R)^{1/2}} =  \frac{1}{(\frac{k^2}{a^2} + m^2)^{1/2}} - \frac{\bar \xi R}{2(\frac{k^2}{a^2} +m^2)^{3/2}} + \mathcal{O}(R^2)  \ . \ee \\
 This simple observation suggests the following ansatz for the first term in the new adiabatic expansion
\be \label{omega0bar} \bar \omega^{(0)}\equiv \bar \omega= \sqrt{\frac{k^2}{a^2(t)}+m^2 + \bar \xi R} \ . \ee
Therefore, the proposed alternative form of the  adiabatic expansion reads [we shall assume  $M^2(t)\equiv m^2 +\bar \xi R >0$]
\be \label{adiabatic2}h_{k}(t) =\frac{1}{\sqrt{\bar W(t)}}e^{-i \int^{t}\bar W(t')dt'} \ , \hspace{0.5cm}  \bar W_k(t) = \bar \omega^{(0)} + \bar \omega^{(1)} + \bar \omega^{(2)} + \cdots \ , \ee
where the function $\bar W_k(t)$ obeys  the differential equation 
\be  \bar W_k^2= \bar \omega^2+\bar \sigma+\frac{3}{4}\frac{\dot{\bar W}_k^2}{\bar W^2_k} -\frac 12\frac{\ddot{\bar W}_k}{\bar W_k}\  \label{equWk3} \ , \ee
with $\bar{\sigma}\equiv\sigma-(\xi-1/6)R$. Having fixed the leading term, the higher-order adiabatic terms are univocally determined. 
Furthermore, we can make use of expressions \eqref{w4}, upgrading $\sigma\to \bar{\sigma}\equiv\sigma-\bar \xi R$ and $\omega\to \bar{\omega}$. It is important to point out that the choice of the leading term  \eqref{omega0bar} does not imply that we consider the function $R$ of adiabatic order zero. This  function is still considered of adiabatic order 2. Hence, if we use \eqref{omegan} to get the adiabatic terms, we must truncate the expressions to fix properly their adiabatic order.
The corresponding expansion for the two-point function up to the $2n$th adiabatic order is  
\bea
\label{PFR} {^{(2n)}}{\bar G}_{PF}(x,x)= \frac{1}{4\pi^2 a^3(t)}\int_0^\infty dk k^2 \left \{ \bar \omega^{-1} + (\bar W^{-1})^{(2)}+ \cdots + (\bar W^{-1})^{(2n)} \right \} \ , 
\eea
where now
\be (\bar W^{-1})^{(2)}=-\frac{5 k^4 \dot{a}^2}{8 a^6 \bar{\omega }^7}+\frac{3 k^2 \dot{a}^2}{4 a^4 \bar{\omega }^5}-\frac{k^2
   \ddot{a}}{4 a^3 \bar{\omega }^5}-\frac{\dot{a}^2}{8 a^2 \bar{\omega }^3}+\frac{\ddot{a}}{4 a
   \bar{\omega}^3} \ . \ee}
We can systematically perform higher-order calculations  assisted, for instance, with the \textit{Mathematica} software.   There is an algorithmic solution to generate recursively  all higher-order terms in the adiabatic expansion. In Appendix \ref{apendixRsumed}, we give more details of this expansion.

\subsection{ Equivalence of the $R$-summed form of the adiabatic expansions} 
Our conjecture is that \eqref{PFR} generates the same expansion as the propagator obtained from the kernel \eqref{DSR} when we restrict to a FLRW spacetime and in the coincident limit, i.e. 
\be\label{conjeture}{^{(2n)}}\bar G_{PF}(x, x)= i {^{(2n)}}\bar G_{SDW}(x, x) \ . \ee
In order to test 
this, we use the above-mentioned result \eqref{equiv1} from Ref. \cite{delRio-Navarro15} and check whether 

\be\label{equivalencen}{^{(2n)}}\bar G_{PF}(x, x)- {^{(2n)}} G_{PF}(x, x)= i \Big({^{(2n)}}\bar G_{SDW}(x, x) - {^{(2n)}}G_{SDW}(x, x)\Big)\ee
holds for a given adiabatic order. Note that \eqref{equivalencen} is equivalent to \eqref{conjeture},  but since both sides of  (\ref{equivalencen}) involve only finite quantities, we can check more directly the proposal. \\ 

The right-hand side in (\ref{equivalencen}) can be written as a finite integral in the proper-time parameter (recall that $m^2\equiv m^2 -i\epsilon$ and this avoids any divergence as $s \to \infty$)
 \be \label{SDWN} {^{(2n)}}\bar G_{SDW}(x, x) - {^{(2n)}}G_{SDW}(x, x)=    \frac{1}{(4\pi)^2}\int_0^\infty \frac{ds}{(is)^2} \sum_{j=0}^n  \left[e^{ -is(m^2 +\bar \xi R)}  \bar a_j (x) (is)^j - e^{-ism^2} a_j (x) (is)^j \right] \ .  \ee
 On the other hand, the left-hand side of \eqref{equivalencen} can be  written as the following integral:
 \bea \label{PFN} {^{(2n)}}\bar G_{PF}(x, x)- {^{(2n)}} G_{PF}(x, x)&=&  \frac{1}{4\pi^2 a^3(t)}\int_0^\infty dk k^2 \sum_{j=0}^n  \left[\left(\bar W^{-1}\right)^{(j)}-\left(W^{-1}\right)^{(j)}\right]. \eea 
 These two integrals are finite by construction,  and can be evaluated analytically.
   First, we will give explicitly the outcome of the integrals above for $n=1,2$ and then we will extend the result for an arbitrary $n$.
  \subsubsection{Cases $n=1, 2$} 
 On one hand, the result of the SDW integral \eqref{SDWN} for $n=1$ is  
  \be \label{resultSDW}{^{(2)}}\bar G_{SDW}(x, x) - {^{(2)}}G_{SDW}(x, x)=  
\frac{-i}{(4\pi)^2}\Big[M^2 \log \left(\frac{M^2}{m^2} \right)-\bar \xi R 
\Big] \ , \ee
where $M^2 = m^2 +\bar \xi R$,  and for  $n=2$, we find
 \be \label{resultSDW}{^{(4)}}\bar G_{SDW}(x, x) - {^{(4)}}G_{SDW}(x, x)=  
\frac{-i}{(4\pi)^2}\Big[M^2 \log \left(\frac{M^2}{m^2} \right)-\bar \xi R  + \Big(\frac{ \bar a_2}{M^{2}}- \frac{a_2}{m^{2}}\Big)\Big]. \ee

 
 On the other hand, one can directly compute the PF integral given in Eq. \eqref{PFN} using the adiabatic expansions given in Secs. \ref{PFadiabaticexp} and \ref{sec3}. For $n=1$, we find 
 \be\label{resultPF}{^{(2)}}\bar G_{PF}(x, x)- {^{(2)}} G_{PF}(x, x)=  \frac{1}{(4\pi)^2}\Big[M^2 \log \left(\frac{M^2}{m^2} \right)-\bar \xi R\big)\Big],\ee 
  and for  $n=2$,
 \be \label{resultPF}{^{(4)}}\bar G_{PF}(x, x) - {^{(4)}}G_{PF}(x, x)=  
\frac{1}{(4\pi)^2}\Big[M^2 \log \left(\frac{M^2}{m^2} \right)-\bar \xi R  +\Big(\frac{ \bar a_2}{M^{2}}- \frac{a_2}{m^{2}}\Big)\Big]. \ee
Comparing \eqref{resultSDW} with \eqref{resultPF}, it is clear that for the lowest adiabatic orders, relation \eqref{equivalencen} is satisfied.
\subsubsection{General case} 
Now, let us generalize the preceding result for an arbitrary $n$. For $n \geq 2$, expression \eqref{SDWN} can be also directly integrated, and the general result is 
 \bea\label{equivalence2b}  i \Big({^{(2n)}}\bar G_{SDW}(x, x) - {^{(2n)}}G_{SDW}(x, x)\Big)=
\frac{1}{(4\pi)^2}\Big[M^2 \log \left(\frac{M^2}{m^2} \right)-\bar \xi R  +\sum_{j=2}^n (j-2)!\Big(\frac{ \bar a_j}{M^{2j-2}}- \frac{a_j}{m^{2j-2}}\Big)\Big].\eea
To evaluate \eqref{PFN}  for $n\geq 2$, one has to compute explicitly the PF adiabatic expansion up to and including the adiabatic order $2n$ and perform the mode integral. Based on 
all previous results, the conjectured result of  ${^{(2n)}}\bar G_{PF}(x, x)- {^{(2n)}} G_{PF}(x, x)$ for $n\geq 2$ is given by
  \bea\label{equivalence2}{^{(2n)}}\bar G_{PF}(x, x)- {^{(2n)}} G_{PF}(x, x)=  i \Big({^{(2n)}}\bar G_{SDW}(x, x) - {^{(2n)}}G_{SDW}(x, x)\Big)=\nonumber\\ 
\frac{1}{(4\pi)^2}\Big[M^2 \log \left(\frac{M^2}{m^2} \right)-\bar \xi R  +\sum_{j=2}^n (j-2)!\Big(\frac{ \bar a_j}{M^{2j-2}}- \frac{a_j}{m^{2j-2}}\Big)\Big].\eea
   We have tested this conjecture up to and including the eight adiabatic order. Since the computations are rather involved, we refer the reader to  Appendix \ref{apC}, where  we give explicit expressions for $\bar a_3$ and $\bar a_4$. We think this provides enough evidence for the general validity of the conjecture. 
 Furthermore, we can  give analytic results for the following finite integrals in both approaches ($n >1$):

\bea
 \frac{1}{4\pi^2 a^3(t)}\int_{0}^{\infty}dkk^2(\bar W^{-1})^{(2n)}=\frac{(n-2)!}{(4\pi)^2}\frac{ \bar a_n}{M^{2n-2}} =  \frac{i}{(4\pi)^2}\int_0^\infty \frac{ds}{(is)^2} e^{ -is(m^2 +\bar \xi R)}  \bar a_n (x) (is)^n \label{barResult} \ .
 \eea

 \section{Generalization: including a scalar background field} \label{sec4}
We can also extend the above discussion to include an external scalar field $\Phi$ with a Yukawa-type coupling. The action of our quantized scalar field $\phi$ is now
\be S_m= \int d^4x\sqrt{-g} \frac{1}{2}( g^{\mu\nu}\nabla_\mu \phi \nabla_\nu \phi -m^2\phi^2 -\xi R\phi^2 -h \Phi \phi^2) \ , \ee 
where $h$ is the coupling constant between both scalar fields. The equation of motion reads
\be (\Box +m^2 +h\Phi+\xi R)\phi=0 \label{motionscalar}\ , \ee
and the expansion for the heat kernel turns out to be 
\be \label{hksE}\langle x, s |  x', 0\rangle =  i\frac{\Delta^{1/2} (x, x')}{(4\pi)^2(is)^2}  e^{\frac{\sigma(x, x')}{2is}} (E_0(x, x') + E_1(x, x') (is) + E_2(x, x') (is)^2 + \cdots ) \ . \ee
 In the coincidence limit $x \to x'$, the  coefficients $E_n$ are 
\bea
E_0(x)=&&1\nonumber\\
E_1(x)=&&-(\bar \xi R+h\Phi)\nonumber\\
E_2(x)=&&\frac{1}{180}R_{\alpha\beta\gamma\delta}R^{\alpha\beta\gamma\delta}-\frac{1}{180}R^{\alpha\beta}R_{\alpha\beta}-\frac{1}{6} \left(\frac15-\xi\right)\Box R+\frac{1}{2}(\bar \xi R + h \Phi )^2+\frac16h \Box \Phi \label{coef}  \ . \ \ \ \ \ 
\eea
The expression for $E_3(x)$ has $46$ terms and it is given, for instance, in Ref. \cite{gilkey}. Note that $\Phi$ should be considered here as a variable of adiabatic order $2$.  The expression of the second coefficient $E_1$ of \eqref{coef} suggests that we can factorize in the same way as before the entire term $\bar \xi R + h\Phi$, also in the form of an exponential.
Hence, we can also write
\be \bar G_{SDW}(x,x) = \frac{1}{(4\pi)^2} \int_0^\infty \frac{ds}{(is)^2} \ e^{ -i(m^2 +\bar \xi  R(x) + h\Phi)s} \sum_j (is)^j \bar E_j (x) \label{hksEbar} \ . \ee 
The lower coefficients in the new expansion are $\bar E_0(x) =1, \bar E_1(x) = 0$ and
\be \bar E_2(x) = \frac{1}{180}R_{\alpha\beta\gamma\delta}R^{\alpha\beta\gamma\delta}-\frac{1}{180}R^{\alpha\beta}R_{\alpha\beta}-\frac{1}{6}\left(\frac15-\xi\right)\Box R+\frac16h \Box \Phi \ . \ee
Note that $\bar E_n$ contain no terms which vanish when $R$ and $\Phi$ (but not their covariant derivatives) are replaced by zero. Therefore, all the dependence on $R$ and $\Phi$ is codified in the exponential in (\ref{hksEbar}). In a similar way, we can redo the Parker-Fulling$-$type adiabatic expansion (\ref{adiabatic2}) with a new choice for the leading term,
\be \label{omega0bar2} \bar \omega^{(0)}\equiv \bar \omega= \sqrt{\frac{k^2}{a^2(t)}+m^2 + \bar \xi  R + h\Phi} \ . \ee
Hence, the results obtained in Sec. \ref{sec3},  regarding the equivalence between both (Schwinger-DeWitt and Parker-Fulling) adiabatic expansions, are now
 \bea\label{resultPFb}{^{(2n)}}\bar G_{PF}(x, x)- {^{(2n)}} G_{PF}(x, x)&=&  \frac{1}{(4\pi)^2}[M^2 \log \left(\frac{M^2}{m^2} \right)-\bar \xi R-h\Phi  +\sum_{j=2}^n (j-2)!\Big(\frac{ \bar E_j}{M^{2j-2}}- \frac{E_j}{m^{2j-2}}\Big)] \nonumber \\
 &=& i\Big({^{(2n)}}\bar G_{SDW}(x, x) - {^{(2n)}}G_{SDW}(x, x)\Big) \ , \eea 
where $M^2$ has been  redefined as $M^2 = m^2 + \bar \xi  R + h\Phi$. \\

We finally remark that one can easily derive approximations for the effective action by simply  using  the summed forms of the adiabatic expansions considered above. Since this is somewhat tangential to the main aim of this paper we refer the reader to Appendix \ref{EA}.


\section{Conclusions}\label{sec5}

The SDW adiabatic expansion of the Feynman propagator is a basic tool in quantum field theory in curved spacetime. A  parallel  WKB-type adiabatic expansion for the field modes in a FLRW spacetime was given by Parker and Fulling in Ref. \cite{parker-fulling74}. Both expansions have been used to implement the renormalization program in curved spacetime and, in particular, in  FLRW universes. \\

The nonperturbative factor $\exp(-is(\xi-\frac{1}{6})R)$ in the heat kernel of SDW expansion, first discovered in Refs. \cite{Parker-Toms85, Jack-Parker}, is of major importance in unraveling  physical consequences  in cosmology \cite{parker-raval,parker-ravalA, parker-vanzella, caldwell-komp-parker-vanzella, Melchiorri}. In this expansion, the focus is not in the renormalization subtractions, which are already well defined in the standard adiabatic expansion. Here, the point is that the nonperturbative factor partially captures nonperturbative effects of the adiabatic vacuum.  
 Within this viewpoint, one could expect that a similar $R$-summed form of the Parker-Fulling adiabatic expansion for the field modes can also be constructed. We have provided here such a construction. We have also tested the equivalence between both Schwinger-DeWitt and Parker-Fulling  $R$-summed expansions in FLRW universes, until and including the adiabatic order 8. We think this provides strong evidence of the equivalence to an arbitrary adiabatic order, as provided by the general formula \eqref{equivalence2}. This can be useful to improve the computations of  physical observables, such as the stress-energy tensor, in the adiabatic approach.  \\

Furthermore, we have added a Yukawa-type interaction between the quantized scalar field and a classical background scalar field $\Phi$, extending the $R$-summed solution to also include a $\Phi$-summed contribution. This is specially relevant in cosmological scenarios where the classical inflation  is coupled to the quantized matter field, as in the  preheating epochs. We believe it could also be interesting to explore additional nonperturbative factorizations for quantized matter field in the presence of  gauge field backgrounds.

\section*{Acknowledgments}

 We thank  P. R. Anderson, P. Beltr\'an-Palau, and A. del Rio for very useful comments and suggestions. This work has been supported by the Spanish MINECO research Grants  No.   FIS2017-84440-C2-1-P and No. FIS2017-91161-EXP and the European Cooperation in Science and Technology (COST) action grant No. CA15117 (CANTATA). A. F. is supported by the Severo Ochoa Ph.D. fellowship SEV-2014-0398-16-1, and the European Social Fund. S. P. is supported by the Formaci\'on del Personal Universitario Ph.D. fellowship FPU16/05287.  Some of the computations have been done with the help of \textit{Mathematica}\texttrademark.

\appendix
\section {Relation between $a_n$ and $\bar  a_n$} \label{Apan}

The relation between the functions $\bar  F(x,x;is)$ and $ F(x,x;is)$ is given by
\bea
\bar  F(x,x;is) \exp(-is\bar \xi R) = F(x,x;is),
\eea
where the functions $\bar  F(x,x;is)$ and $ F(x,x;is)$ can be expanded  in powers of $s$ as
\bea
\bar F(x,x;is) =\bar  a_0(x)+\bar  a_1(x)(is)+\bar  a_2(x)(is)^2+...\\
F(x,x;is) = a_0(x)+a_1(x)(is)+a_2(x)(is)^2+...
\eea
Expanding the exponential in powers of the scalar curvature and combining the terms with equal powers of $s$, we arrive to the following relation between $a_n$ and $\bar  a_n$:
\bea
\bar a_n=\sum_{k=0}^{n} a_{n-k}\frac{(\bar \xi R)^{k}}{k!} \label{relhat}
\eea
In particular, for the first terms, we have
\bea
\bar a_0 &=&a_0=1,\\
\bar a_1 &=&a_1+\bar \xi R=0,\\
\bar a_2 &=&a_2-\frac12(\bar \xi R)^2,\\
\bar a_3 &=&
a_3 + a_2\bar \xi R-\frac{1}{3}(\bar \xi R)^3,
\eea
where we have used $a_0=1$ and $a_1=-\bar \xi R$. If we add an external background field, relation \eqref{relhat} also holds for the coefficients $E_n$ by doing the change $\bar \xi R\to \bar \xi R+h\Phi$.
%

\section {$R$-summed Parker-Fulling adiabatic expansion } \label{apendixRsumed}

In this section, we will briefly explain some details on the $R$-summed Parker-Fulling adiabatic expansion introduced in Sec. \ref{sec3}. Starting from the mode equation for the scalar field $\ddot h_k (t)+(\omega^2+\sigma)h_k(t)=0$, and proposing the usual WKB ansatz
\be \label{AnsatzApendix}h_{k}(t) =\frac{1}{\sqrt{\bar W(t)}}e^{-i \int^{t}\bar W(t')dt'} \ , \hspace{0.5cm}  \bar W_k(t) = \bar \omega^{(0)} + \bar \omega^{(1)} + \bar \omega^{(2)} + \cdots \ , \ee
 we arrive to the following equation for $\bar W_k(t)$:
 \be \bar W_k^2= \omega^2+\sigma+\frac{3}{4}\frac{\dot{\bar W}_k^2}{\bar W^2_k} -\frac 12\frac{\ddot{\bar W}_k}{\bar W_k}  \label{equWk3A} \ . \ee
 
 If we fix the leading term of the expansion as 
 \be \label{omega0A} \bar \omega^{(0)}\equiv \bar \omega=\sqrt{\omega^2 + \bar \xi R}= \sqrt{\frac{k^2}{a^2}+m^2  + \bar \xi R} \ , \ee
Eq. \eqref{equWk3A} can be rewritten as 
 \be  \bar W_k^2= \bar \omega^2+\bar \sigma+\frac{3}{4}\frac{\dot{\bar W}_k^2}{\bar W^2_k} -\frac 12\frac{\ddot{\bar W}_k}{\bar W_k} \ , \ee
 where $\bar \sigma=\sigma-\bar \xi R$, and we can obtain the terms of the adiabatic expansion \eqref{AnsatzApendix} as usual: expanding the function $\bar W_k$ adiabatically and 
and regrouping all terms with the same adiabatic order.
\\
 
 Note that the choice of the leading term as in \eqref{omega0A} does not imply that we consider the function $R$ of adiabatic order 0. This  function is still considered of adiabatic order 2. It means that the time derivative of the leading order
  \bea
 \dot {\bar \omega} =-\frac{2k^2 \dot a}{2a^3\bar \omega}+\frac{\bar \xi \dot R }{2\bar \omega}
 \eea
 will contain terms of adiabatic order 1 ($\sim \dot a$), but also of adiabatic order 3 ($\sim \dot R$). As a consequence, if we use the formulas given in \eqref{w4} to compute the next-to-leading-order terms of the adiabatic expansion, namely $\bar \omega^{(2)}$, $\bar \omega^{(2)}$..., we have to truncate the resulting expressions to get only the terms with the correct adiabatic order. For example, for $\bar \omega ^{(2)}$  we get
\bea
\bar \omega ^{(2)}=\frac{\bar \sigma}{2 \bar \omega} +\frac{3\dot {\bar \omega} ^2}{8 \bar \omega^3}-\frac{\ddot \omega}{4 \bar \omega ^2}\bigg\rvert_{(2)}=+\frac{\bar{\sigma }}{2 \bar{\omega }}+\frac{5 \dot{a}^2 k^4}{8 a^6 \bar{\omega }^5}-\frac{3 \dot{a}^2 k^2}{4 a^4 \bar{\omega }^3}+\frac{k^2 \ddot{a}}{4 a^3 \bar{\omega }^3}.
\eea
 Similarly, for $\bar \omega ^{(4)}$ and $\bar \omega ^{(6)}$, we obtain

\bea
 \bar \omega^{(4)}&=&-\frac{1105 \dot{a}^4 k^8}{128 a^{12} \bar{\omega }^{11}}+\frac{663 \dot{a}^4 k^6}{32 a^{10} \bar{\omega }^9}-\frac{221 \dot{a}^2 k^6 \ddot{a}}{32 a^9 \bar{\omega }^9}-\frac{507 \dot{a}^4
   k^4}{32 a^8 \bar{\omega }^7}+\frac{183 \dot{a}^2 k^4 \ddot{a}}{16 a^7 \bar{\omega }^7}-\frac{25 \dot{a}^2 k^4 \bar{\sigma }}{16 a^6 \bar{\omega }^7}-\frac{19 k^4 \ddot{a}^2}{32 a^6
   \bar{\omega }^7}\nonumber \\
   &\, \,\,&  +\frac{15 \dot{a}^4 k^2}{4 a^6 \bar{\omega }^5}-\frac{9 \dot{a}^2 k^2 \ddot{a}}{2 a^5 \bar{\omega }^5}+\frac{9 \dot{a}^2 k^2 \bar{\sigma }}{8 a^4 \bar{\omega }^5}+\frac{9
   k^2 \ddot{a}^2}{16 a^4 \bar{\omega }^5}-\frac{5 \dot{a} \dot{g} k^2}{8 a^3 \bar{\omega }^5}-\frac{3 k^2 \bar{\sigma } \ddot{a}}{8 a^3 \bar{\omega }^5}-\frac{5 \dot{a} k^2
   \dot{\bar{\sigma }}}{8 a^3 \bar{\omega }^5}\nonumber \\
   &\, \,\,&  -\frac{7 \dot{a} a^{(3)} k^4}{8 a^6 \bar{\omega }^7}+\frac{3 \dot{a} a^{(3)} k^2}{4 a^4 \bar{\omega }^5}-\frac{a^{(4)} k^2}{16 a^3 \bar{\omega
   }^5}-\frac{\ddot{g}}{8 \bar{\omega }^3}-\frac{\ddot{\bar{\sigma }}}{8 \bar{\omega }^3}-\frac{\bar{\sigma }^2}{8 \bar{\omega }^3},
\eea
where $g(t)=\bar \xi R$, and
\bea
\bar \omega^{(6)}&=&\frac{414125 \dot{a}^6 k^{12}}{1024 a^{18} \bar{\omega }^{17}}-\frac{745425 \dot{a}^6 k^{10}}{512 a^{16} \bar{\omega }^{15}}+\frac{248475 \dot{a}^4 \ddot{a} k^{10}}{512 a^{15} \bar{\omega
   }^{15}}+\frac{513087 \dot{a}^6 k^8}{256 a^{14} \bar{\omega }^{13}}+\frac{12155 \bar{\sigma } \dot{a}^4 k^8}{256 a^{12} \bar{\omega }^{13}}+\frac{34503 \dot{a}^2 \ddot{a}^2 k^8}{256
   a^{12} \bar{\omega }^{13}}\nonumber \\
   &\, \,\,& -\frac{179469 \dot{a}^4 \ddot{a} k^8}{128 a^{13} \bar{\omega }^{13}}+\frac{1055 \dot{a}^3 a^{(3)} k^8}{16 a^{12} \bar{\omega }^{13}}-\frac{166089 \dot{a}^6
   k^6}{128 a^{12} \bar{\omega }^{11}}-\frac{5967 \bar{\sigma } \dot{a}^4 k^6}{64 a^{10} \bar{\omega }^{11}}+\frac{631 \ddot{a}^3 k^6}{128 a^9 \bar{\omega }^{11}}-\frac{9513 \dot{a}^2
   \ddot{a}^2 k^6}{32 a^{10} \bar{\omega }^{11}}\nonumber \\
   &\, \,\,& +\frac{184215 \dot{a}^4 \ddot{a} k^6}{128 a^{11} \bar{\omega }^{11}}+\frac{1989 \bar{\sigma } \dot{a}^2 \ddot{a} k^6}{64 a^9 \bar{\omega
   }^{11}}+\frac{1105 \dot{a}^3 \dot{g} k^6}{64 a^9 \bar{\omega }^{11}}+\frac{1105 \dot{a}^3 \dot{\bar{\sigma }} k^6}{64 a^9 \bar{\omega }^{11}}-\frac{9063 \dot{a}^3 a^{(3)} k^6}{64 a^{10}
   \bar{\omega }^{11}}+\frac{1391 \dot{a} \ddot{a} a^{(3)} k^6}{64 a^9 \bar{\omega }^{11}}\nonumber \\
   &\, \,\,& +\frac{815 \dot{a}^2 a^{(4)} k^6}{128 a^9 \bar{\omega }^{11}}+\frac{6147 \dot{a}^6 k^4}{16 a^{10}
   \bar{\omega }^9}+\frac{3549 \bar{\sigma } \dot{a}^4 k^4}{64 a^8 \bar{\omega }^9}-\frac{495 \ddot{a}^3 k^4}{64 a^7 \bar{\omega }^9}+\frac{175 \bar{\sigma }^2 \dot{a}^2 k^4}{64 a^6
   \bar{\omega }^9}+\frac{26199 \dot{a}^2 \ddot{a}^2 k^4}{128 a^8 \bar{\omega }^9}\nonumber \\
   &\, \,\,& +\frac{133 \bar{\sigma } \ddot{a}^2 k^4}{64 a^6 \bar{\omega }^9}+\frac{69 \left(a^{(3)}\right)^2 k^4}{128
   a^6 \bar{\omega }^9}-\frac{2427 \dot{a}^4 \ddot{a} k^4}{4 a^9 \bar{\omega }^9}-\frac{1281 \bar{\sigma } \dot{a}^2 \ddot{a} k^4}{32 a^7 \bar{\omega }^9}-\frac{663 \dot{a}^3 \dot{g}
   k^4}{32 a^7 \bar{\omega }^9}+\frac{221 \dot{a} \ddot{a} \dot{g} k^4}{32 a^6 \bar{\omega }^9}\nonumber \\
   &\, \,\,& +\frac{221 \dot{a}^2 \ddot{g} k^4}{64 a^6 \bar{\omega }^9}-\frac{663 \dot{a}^3
   \dot{\bar{\sigma }} k^4}{32 a^7 \bar{\omega }^9}+\frac{221 \dot{a} \ddot{a} \dot{\bar{\sigma }} k^4}{32 a^6 \bar{\omega }^9}+\frac{221 \dot{a}^2 \ddot{\bar{\sigma }} k^4}{64 a^6
   \bar{\omega }^9}+\frac{189 \dot{a}^3 a^{(3)} k^4}{2 a^8 \bar{\omega }^9}+\frac{49 \bar{\sigma } \dot{a} a^{(3)} k^4}{16 a^6 \bar{\omega }^9}\nonumber \\
   &\, \,\,& -\frac{2091 \dot{a} \ddot{a} a^{(3)} k^4}{64
   a^7 \bar{\omega }^9}-\frac{285 \dot{a}^2 a^{(4)} k^4}{32 a^7 \bar{\omega }^9}+\frac{55 \ddot{a} a^{(4)} k^4}{64 a^6 \bar{\omega }^9}+\frac{27 \dot{a} a^{(5)} k^4}{64 a^6 \bar{\omega
   }^9}-\frac{315 \dot{a}^6 k^2}{8 a^8 \bar{\omega }^7} -\frac{75 \bar{\sigma } \dot{a}^4 k^2}{8 a^6 \bar{\omega }^7}\nonumber \\
   &\, \,\,& +\frac{45 \ddot{a}^3 k^2}{16 a^5 \bar{\omega }^7}-\frac{45 \bar{\sigma
   }^2 \dot{a}^2 k^2}{32 a^4 \bar{\omega }^7}-\frac{675 \dot{a}^2 \ddot{a}^2 k^2}{16 a^6 \bar{\omega }^7}-\frac{45 \bar{\sigma } \ddot{a}^2 k^2}{32 a^4 \bar{\omega }^7}-\frac{15
   \left(a^{(3)}\right)^2 k^2}{32 a^4 \bar{\omega }^7}+\frac{675 \dot{a}^4 \ddot{a} k^2}{8 a^7 \bar{\omega }^7}+\frac{45 \bar{\sigma } \dot{a}^2 \ddot{a} k^2}{4 a^5 \bar{\omega
   }^7}\nonumber \\
   &\, \,\,& +\frac{15 \bar{\sigma }^2 \ddot{a} k^2}{32 a^3 \bar{\omega }^7}+\frac{21 \dot{a}^3 \dot{g} k^2}{4 a^5 \bar{\omega }^7}+\frac{25 \bar{\sigma } \dot{a} \dot{g} k^2}{16 a^3 \bar{\omega
   }^7}-\frac{63 \dot{a} \ddot{a} \dot{g} k^2}{16 a^4 \bar{\omega }^7}-\frac{57 \dot{a}^2 \ddot{g} k^2}{32 a^4 \bar{\omega }^7}+\frac{19 \ddot{a} \ddot{g} k^2}{32 a^3 \bar{\omega
   }^7}+\frac{21 \dot{a}^3 \dot{\bar{\sigma }} k^2}{4 a^5 \bar{\omega }^7}+\frac{25 \bar{\sigma } \dot{a} \dot{\bar{\sigma }} k^2}{16 a^3 \bar{\omega }^7}\nonumber \\
   &\, \,\,& -\frac{63 \dot{a} \ddot{a}
   \dot{\bar{\sigma }} k^2}{16 a^4 \bar{\omega }^7}-\frac{57 \dot{a}^2 \ddot{\bar{\sigma }} k^2}{32 a^4 \bar{\omega }^7}+\frac{19 \ddot{a} \ddot{\bar{\sigma }} k^2}{32 a^3 \bar{\omega
   }^7}-\frac{75 \dot{a}^3 a^{(3)} k^2}{4 a^6 \bar{\omega }^7}-\frac{15 \bar{\sigma } \dot{a} a^{(3)} k^2}{8 a^4 \bar{\omega }^7}+\frac{45 \dot{a} \ddot{a} a^{(3)} k^2}{4 a^5 \bar{\omega
   }^7}+\frac{7 \dot{g} a^{(3)} k^2}{16 a^3 \bar{\omega }^7}\nonumber \\
   &\, \,\,& +\frac{7 \dot{\bar{\sigma }} a^{(3)} k^2}{16 a^3 \bar{\omega }^7}+\frac{7 \dot{a} g^{(3)} k^2}{16 a^3 \bar{\omega }^7}+\frac{7
   \dot{a} \bar{\sigma }^{(3)} k^2}{16 a^3 \bar{\omega }^7}+\frac{45 \dot{a}^2 a^{(4)} k^2}{16 a^5 \bar{\omega }^7}-\frac{45 \ddot{a} a^{(4)} k^2}{64 a^4 \bar{\omega }^7}+\frac{5
   \bar{\sigma } a^{(4)} k^2}{32 a^3 \bar{\omega }^7} -\frac{9 \dot{a} a^{(5)} k^2}{32 a^4 \bar{\omega }^7}\nonumber \\
   &\, \,\,& +\frac{a^{(6)} k^2}{64 a^3 \bar{\omega }^7}+\frac{5 \dot{g}^2}{32 \bar{\omega
   }^5}+\frac{5 \dot{\bar{\sigma }}^2}{32 \bar{\omega }^5}+\frac{3 \bar{\sigma } \ddot{g}}{16 \bar{\omega }^5}+\frac{5 \dot{g} \dot{\bar{\sigma }}}{16 \bar{\omega }^5}+\frac{3 \bar{\sigma }
   \ddot{\bar{\sigma }}}{16 \bar{\omega }^5}+\frac{g^{(4)}}{32 \bar{\omega }^5}+\frac{\bar{\sigma }^{(4)}}{32 \bar{\omega }^5}+\frac{\bar{\sigma }^3}{16 \bar{\omega }^5}
.\eea \\

Note that odd adiabatic terms are zero, $\omega^{(2r+1)}=0$. The function $(\bar W_k)^{-1}$ is also expanded adiabatically,
\bea
(\bar W_k)^{-1}=\bar \omega^{-1}+(\bar W^{-1})^{(2)}+(\bar W^{-1})^{(4)}...
\eea
and the first terms of the expansion are
\bea
(\bar W^{-1})^{(2)}=-\frac{\bar \omega ^{(2)}}{\bar \omega ^2},\hspace{0.6cm} (\bar W^{-1})^{(4)}=-\frac{\bar \omega ^{(4)}}{\bar \omega ^2}+\frac{(\bar \omega ^{(2)})^2}{\bar \omega ^3}
\ . \eea
The results above are also valid when we add an scalar field background $h\Phi$ by upgrading $\bar \omega \to \sqrt{\omega^2+\bar \xi R+h\Phi}$ and $g(t) \to \bar \xi R + h \Phi$.

\section {$\bar a_3$ and $\bar a_4$ coefficients} \label{apC}
 In this Appendix we give the $R$-summed  coefficients of adiabatic orders 6 and 8, namely $\bar a_3$ and $\bar a_4$ for a FLRW metric:
 
 \bea \label{a3}
\bar  a_3&=& -\frac{12 \dot{a}^6 \xi ^2}{a^6}+\frac{4 \dot{a}^6 \xi }{a^6}-\frac{a^{(6)} \xi }{10
   a}-\frac{\dot{a}^6}{3 a^6}+\frac{3 a^{(6)}}{140 a}+\frac{12 \dot{a}^4 \xi ^2
   \ddot{a}}{a^5}-\frac{21 \dot{a}^4 \xi  \ddot{a}}{10 a^5}-\frac{11 \dot{a}^4
   \ddot{a}}{420 a^5}-\frac{3 \dot{a}^2 \xi ^2 \ddot{a}^2}{a^4}\nonumber \\&-&\frac{47 \dot{a}^2 \xi
    \ddot{a}^2}{10 a^4}+\frac{109 \dot{a}^2 \ddot{a}^2}{105 a^4}-\frac{6 a^{(3)}
   \dot{a} \xi ^2 \ddot{a}}{a^3}+\frac{67 a^{(3)} \dot{a} \xi  \ddot{a}}{10
   a^3}+\frac{7 \xi  \ddot{a}^3}{5 a^3}-\frac{67 a^{(3)} \dot{a} \ddot{a}}{60
   a^3}-\frac{5 \ddot{a}^3}{18 a^3}-\frac{2 a^{(5)} \dot{a} \xi }{5 a^2}\nonumber  \\&+&\frac{3
   a^{(5)} \dot{a}}{35 a^2}+\frac{12 a^{(3)} \dot{a}^3 \xi ^2}{a^4}-\frac{59 a^{(3)}
   \dot{a}^3 \xi }{10 a^4}+\frac{9 a^{(4)} \dot{a}^2 \xi }{10 a^3}+\frac{97 a^{(3)}
   \dot{a}^3}{140 a^4}-\frac{11 a^{(4)} \dot{a}^2}{60 a^3}-\frac{a^{(4)} \xi 
   \ddot{a}}{2 a^2}\nonumber  \\&+&\frac{41 a^{(4)} \ddot{a}}{420 a^2}-\frac{3 \left(a^{(3)}\right)^2
   \xi ^2}{a^2}+\frac{7 \left(a^{(3)}\right)^2 \xi }{10
   a^2}-\frac{\left(a^{(3)}\right)^2}{42 a^2}
   \eea

 \   \bea
  \bar a_4&=& \frac{96 \xi ^2 \dot{a}^8}{5 a^8}-\frac{32 \xi  \dot{a}^8}{5 a^8}+\frac{8
   \dot{a}^8}{15 a^8}-\frac{432 \xi ^2 \ddot{a} \dot{a}^6}{5 a^7}+\frac{3807 \xi 
   \ddot{a} \dot{a}^6}{140 a^7}-\frac{291 \ddot{a} \dot{a}^6}{140 a^7}+\frac{108 \xi
   ^2 a^{(3)} \dot{a}^5}{5 a^6}-\frac{783 \xi  a^{(3)} \dot{a}^5}{140 a^6}\nonumber \\&+&\frac{39
   a^{(3)} \dot{a}^5}{140 a^6}+\frac{981 \xi ^2 \ddot{a}^2 \dot{a}^4}{10
   a^6}-\frac{493 \xi  \ddot{a}^2 \dot{a}^4}{20 a^6}+\frac{937 \ddot{a}^2
   \dot{a}^4}{840 a^6}-\frac{19 \xi  a^{(4)} \dot{a}^4}{20 a^5}+\frac{79 a^{(4)}
   \dot{a}^4}{420 a^5}-\frac{63 \xi ^2 \ddot{a} a^{(3)} \dot{a}^3}{5 a^5}\nonumber  \\&-&\frac{209
   \xi  \ddot{a} a^{(3)} \dot{a}^3}{35 a^5}+\frac{71 \ddot{a} a^{(3)} \dot{a}^3}{42
   a^5}-\frac{12 \xi ^2 a^{(5)} \dot{a}^3}{5 a^4}+\frac{17 \xi  a^{(5)} \dot{a}^3}{14
   a^4}-\frac{31 a^{(5)} \dot{a}^3}{210 a^4}-\frac{138 \xi ^2 \ddot{a}^3 \dot{a}^2}{5
   a^5}-\frac{13 \xi  \ddot{a}^3 \dot{a}^2}{35 a^5}+\frac{53 \ddot{a}^3 \dot{a}^2}{45
   a^5}\nonumber  \\&-&\frac{15 \xi ^2 \left(a^{(3)}\right)^2 \dot{a}^2}{2 a^4}  +\frac{361 \xi 
   \left(a^{(3)}\right)^2 \dot{a}^2}{70 a^4}-\frac{209 \left(a^{(3)}\right)^2
   \dot{a}^2}{280 a^4}-\frac{9 \xi ^2 \ddot{a} a^{(4)} \dot{a}^2}{a^4}+\frac{69 \xi 
   \ddot{a} a^{(4)} \dot{a}^2}{10 a^4}-\frac{73 \ddot{a} a^{(4)} \dot{a}^2}{70
   a^4}-\frac{\xi  a^{(6)} \dot{a}^2}{10 a^3}\nonumber  \\&+&\frac{3 a^{(6)} \dot{a}^2}{140
   a^3}-\frac{9 \xi ^2 \ddot{a}^2 a^{(3)} \dot{a}}{a^4}+\frac{90 \xi  \ddot{a}^2
   a^{(3)} \dot{a}}{7 a^4}-\frac{95 \ddot{a}^2 a^{(3)} \dot{a}}{42 a^4}+\frac{27 \xi
   ^2 a^{(3)} a^{(4)} \dot{a}}{5 a^3}-\frac{237 \xi  a^{(3)} a^{(4)} \dot{a}}{70
   a^3}+\frac{199 a^{(3)} a^{(4)} \dot{a}}{420 a^3}\nonumber  \\&+&\frac{6 \xi ^2 \ddot{a} a^{(5)}
   \dot{a}}{5 a^3}-\frac{43 \xi  \ddot{a} a^{(5)} \dot{a}}{35 a^3}+\frac{29 \ddot{a}
   a^{(5)} \dot{a}}{140 a^3}+\frac{\xi  a^{(7)} \dot{a}}{28 a^2}-\frac{a^{(7)}
   \dot{a}}{126 a^2}+\frac{9 \xi ^2 \ddot{a}^4}{10 a^4}+\frac{17 \xi  \ddot{a}^4}{14
   a^4}-\frac{713 \ddot{a}^4}{2520 a^4}+\frac{21 \xi ^2 \ddot{a}
   \left(a^{(3)}\right)^2}{5 a^3}\nonumber  \\&-&\frac{95 \xi  \ddot{a} \left(a^{(3)}\right)^2}{28
   a^3}+\frac{18 \ddot{a} \left(a^{(3)}\right)^2}{35 a^3}+\frac{9 \xi ^2
   \left(a^{(4)}\right)^2}{10 a^2}-\frac{39 \xi  \left(a^{(4)}\right)^2}{140
   a^2}+\frac{11 \left(a^{(4)}\right)^2}{504 a^2}+\frac{9 \xi ^2 \ddot{a}^2 a^{(4)}}{5
   a^3}-\frac{69 \xi  \ddot{a}^2 a^{(4)}}{35 a^3}\nonumber  \\&+&\frac{137 \ddot{a}^2 a^{(4)}}{420
   a^3}+\frac{6 \xi ^2 a^{(3)} a^{(5)}}{5 a^2}-\frac{12 \xi  a^{(3)} a^{(5)}}{35
   a^2}+\frac{29 a^{(3)} a^{(5)}}{1260 a^2}+\frac{2 \xi  \ddot{a} a^{(6)}}{35
   a^2}-\frac{\ddot{a} a^{(6)}}{90 a^2}+\frac{\xi  a^{(8)}}{140 a}-\frac{a^{(8)}}{630
   a}.
   \eea

 \section {Effective action} \label{EA}

We find it useful to show how to derive  approximate one-loop effective actions with the adiabatic expansions introduced in the body of the paper.   The formal quantum effective action $W$, obtained by integrating out the degrees of freedom of the quantized scalar field, is given by \cite{parker-toms}
\bea W= S_{class}- \frac{1}{2}\int d^4x \int_0^\infty \frac{ds}{is} e^{-ism^2} \langle x, s | x, 0 \rangle  \ , \eea
 where $S_{class}$ is the classical action including the Yukawa coupling and the gravitational interaction. The quantum corrected part can be further written as 
\bea  \frac{1}{32\pi^2}\int d^4x \int^{\infty}_0\frac{ds}{s^3}e^{-ism^2}F(x,x;is)  \ . \eea
The expression above is UV divergent, and it requires renormalization subtractions up to and including the fourth adiabatic order.  Following the same approach as for the Feynman propagator, we will use the extended $R$-summed expansion until second adiabatic order to approximate $F(x,x;is)$  and the usual SDW expansion until second order for the subtraction terms. This ensures that the final quantity is finite. We have then
\be  W= S_{class}+\frac{1}{32\pi^2}\int d^4x \int^{\infty}_0\frac{ds}{s^3}e^{-ism^2}[e^{-is\left(\bar \xi R + h\Phi\right) }\left(1+\bar{E}_1(x)(is)+ \bar{E}_2(x)(is)^2\right)- \left(1+ E_1(x)(is)+ E_2(x)(is)^2\right)]  \ . \ee
 With this input, and after performing the {\it finite} integration in proper-time $ds$, one gets 
\be W = S_{class} +  \int d^4x \ L_{eff} \ee
 where 
\bea
&&L_{eff}=\frac{1}{64\pi^2}\left\{(\bar \xi R+h\Phi)\left(m^2+\frac32\left(\bar \xi R+h\Phi\right)\right)-\left(M^4+2\bar{E}_2(x)\right)\log{ \left |\frac{M^2}{m^2}\right |}\right\} \nonumber \\&&
+\frac{i}{64\pi}\left[M^4+2\bar{E_2}\right]\Theta(-M^2)
\ . \eea 
The imaginary part of the effective action  accounts for the particle creation phenomena induced by the given metric and also by the scalar field background.
Note that, for $h=0$ (no coupling to the scalar background field), we recover the effective action calculated in Ref. \cite{parker-raval} by means of the $\zeta$-function regularization. Here, we have derived the effective action within the adiabatic approach and in a very straightforward way.  \\

\end{document}